# High-temperature helical edge states in BiSbTeSe$_2$/graphene van der Waals heterostructure

Yoichi Tanabe[1*†], Ngoc Han Tu[2*†], Ming-Chun Jiang[2,3], Yi Ling Chiew[2], Mitsutaka Haruta[4], Kiyohiro Adachi[2], David Pomaranski[5], Ryo Ito[2], Yuya Shimazaki[2,5], Daisuke Hashizume[2], Xiuzhen Yu[2], Guang-Yu Guo[3,6], Ryotaro Arita[2,7], and Michihisa Yamamoto[2,5*]

[1] Department of Applied Science, Faculty of Science, Okayama University of Science, Japan
[2] Center for Emergent Matter Science (CEMS), RIKEN, Saitama, Japan
[3] Department of Physics and Center for Theoretical Physics, National Taiwan University, Taiwan
[4] Institute for Chemical Research (ICR), Kyoto University, Kyoto, Japan
[5] Quantum-Phase Electronics Center and Department of Applied Physics, The University of Tokyo, Tokyo, Japan
[6] Physics Division, National Center for Theoretical Sciences, Taipei 10617, Taiwan
[7] Department of Physics, The University of Tokyo, Tokyo, Japan
* tanabe@ous.ac.jp, han.tu@riken.jp, michihisa.yamamoto@riken.jp
† These authors have equal contributions.

**Van der Waals heterostructures have been used to tailor atomic layers into various artificial materials through interactions at heterointerfaces. The interplay between the band gap created by the band folding of the interfacial potential and the band inversion driven by enhanced spin-orbit interaction (SOI) through band hybridization enables us to realize a two-dimensional topological insulator (2D-TI). Here we report the realization of graphene 2D-TIs by epitaxial growth of three-dimensional topological insulator (3D-TI) BiSbTeSe$_2$ ultrathin films on graphene. By increasing the BiSbTeSe$_2$ thickness from 2 nm to 9 nm to enhance SOI on graphene, the electronic state is altered from the trivial Kekulé insulator to the 2D-TI. The nonlocal transport reveals the helical edge conduction which survives up**

**to 200 K at maximum. Our graphene 2D-TI is stable, easy to make electrical contacts, and of high quality. It offers various applications including spin-current conversion and platforms for Majorana fermions in junctions to superconductors.**

Main

Van der Waals (vdW) heterostructures, created by stacking two-dimensional (2D) atomic layers, offer a novel approach for manipulating device properties through interfacial interactions between the layers [1-4]. The combination of graphene and topological materials is also anticipated to generate new topological materials by utilizing interfacial lattice matching for the reconstruction of the electronic structure and the enhancement of the spin-orbit interaction (SOI) through band hybridization [5-8]. We consider a theoretical proposal involving van der Waals (vdW) heterostructures composed of graphene and thin films of three-dimensional topological insulators (3D-TIs) (Fig. 1a). In such structures, aligning the Dirac neutral point (DP) of graphene and the 3D-TI is theoretically expected to significantly enhance the intrinsic spin–orbit interaction (SOI) in graphene. This enhancement could enable the emergence of graphene-based topological insulator (2D-TI) states at elevated temperatures. Graphene, the first theoretically predicted 2D-TI [9,10], has so far faced challenges in realizing 2D-TI states at practical temperatures due to its inherently weak SOI. In contrast, tetradymite compounds exhibit 3D-TI states driven by the strong SOI [11-14]. A number of theoretical studies have suggested that integrating graphene with thin 3D-TI films in a vdW heterostructure can induce high-temperature 2D-TI states, enabled by proximity-induced enhancement of SOI in graphene [6,15]. One of the most promising approaches involves the epitaxial growth of an ultrathin 3D-TI tetradymite compound directly on graphene. When the 3D-TI has a lattice constant that matches the $\sqrt{3}a \times \sqrt{3}a$ supercell of graphene (where $a$ is graphene lattice constant), a Kekulé distortion is induced at the heterostructure interface. This distortion folds graphene's massless Dirac cones at the *K* and *K'* points into the *Γ* point, and opens a band gap [16-20]. Meanwhile, in ultrathin 3D-TI films, a hybridization gap emerges at the *Γ* point due to coupling between the top and bottom surface states (Fig. 1c) [21,22]. When the DPs of graphene and the 3D-TI film are properly aligned, their proximity near the *Γ* point can significantly enhance the intrinsic SOI in graphene. The effective SOI in this hybrid system can be expressed as: $\lambda_{SO} = \Lambda_{SO} \frac{|t_1|^2}{2\sqrt{3}(\varepsilon_{TI}-\varepsilon_G)^2}$ [6], where $\Lambda_{SO}$ is the intrinsic SOI of 3D-TI, $t_1$ is the hopping parameter between TI surface states and graphene, $\varepsilon_{TI}, \varepsilon_G$ are the energies of the states near to the DP of graphene and 3D-TI, respectively. While the strength of $\Lambda_{SO}$ can be tuned by adjusting the thickness of the 3D-TI films, a close alignment of the DP energies ($\varepsilon_{TI} \sim \varepsilon_G$) leads to a resonance-like enhancement of $\lambda_{SO}$, potentially driving the system from a Kekulé insulator to a 2D-TI state [6]. Therefore, exploring the thickness dependence of 3D-TI films, along with precise control of DP alignment is crucial for identifying the optimal conditions that maximize $\lambda_{SO}$ while maintaining the hybridization gap of the 3D-TI - an essential factor for realizing robust 2D-TI states at elevated temperatures. However, graphene-based 2D-TIs in van der Waals (vdW) heterostructures have yet to be experimentally realized. This is primarily due to the energy misalignment between the DPs of graphene and those of 3D-TIs, which arise from the significant bulk carrier contributions present in tetradymite-type 3D-TI compounds. While many theoretical candidates for 2D-TIs have been proposed, only a few have been experimentally confirmed through electrical transport measurements [23-31]. Quantum well structures created by molecular beam epitaxy (MBE) exhibit small band gaps, leading to observable effects only at low temperatures. The monolayer of $WTe_2$, one of the most promising 2D-TIs to date, faces challenges in practical applications owing to the environmental sensitivity and sample size, which limits its device manufacturability. In the case of graphene, decorating its surface with dilute $Bi_2Te_3$ nanoparticles - one of the tetradymite compounds - has successfully converted graphene into a 2D-TI state with an inversion gap at the $K$ and $K'$ points [31]. This technique, however, suffers from low and uneven coverage, as well as the influence of $Bi_2Te_3$ bulk carriers, making it difficult to control and reproduce.

      We have developed a fabrication technique for constructing van der Waals (vdW) heterostructures composed of graphene and ultrathin $BiSbTeSe_2$ films - one of the most ideal 3D-TIs among tetradymite compounds, known for its excellent bulk insulating properties [32–35]. By employing ultrathin $BiSbTeSe_2$ films - where a hybridization gap forms in the 3D-TI surface states around the $\Gamma$ point for thicknesses ≤ 9 nm [22] - we investigate the alignment of their DP energies, enabling control over the Fermi level near the DPs. $BiSbTeSe_2$ features a layered structure with five covalently bonded atomic layers forming

a quintuple layer (QL) (Fig. 1a). The epitaxial growth of BiSbTeSe$_2$ on monolayer graphene, despite a slight lattice mismatch between its in-plane lattice constant (approximately 4.18 Å) and the $\sqrt{3}\,a \times \sqrt{3}\,a$ supercell of graphene (~ 4.23 Å), is expected to induce a Kekulé distortion in the graphene layer (Figs. 1 a, b). The combination of the Kekulé gap in graphene and the hybridization gap in the 3D-TI tends to drive the system into a trivial insulating state for BiSbTeSe$_2$ film thicknesses up to 9 nm. However, as the film thickness increases from 2 nm to 9 nm, electrical transport measurements indicate a transition from a trivial insulator to a 2D -TI state, likely driven by the enhancement of the effective spin–orbit interaction (SOI). Notably, the temperature window for observing helical edge conduction narrows with increasing film thickness - from 3 nm to 9 nm - due to competition between the SOI-induced band inversion in graphene and the hybridization gap in BiSbTeSe$_2$. In thicker samples (~ 20 nm), where the hybridization gap is fully suppressed, graphene serves as an excellent insulating substrate for the epitaxial growth of high-quality BiSbTeSe$_2$ as evidenced by the observation of $\nu = 1$ quantum Hall states [extended Data Fig. 7]. Our results demonstrate that ultrathin BiSbTeSe$_2$/graphene van der Waals heterostructures offer a versatile platform for realizing and tuning novel topological phases.

**Fabrication of BiSbTeSe$_2$/graphene van der Waals heterostructures**

vdW heterostructures of graphene and BiSbTeSe$_2$ thin films were fabricated by depositing BiSbTeSe$_2$ films on the surface of commercially available CVD graphene via physical vapor deposition method [35]. The details of growth set up are shown in the Method section and the Extended Data Fig. 1. Figure 1a presents a schematic of our proposed growth mode for BiSbTeSe$_2$ films on graphene, based on their lattice constants. The atomic force microscopy (AFM) image in Fig. 1f shows that the variation of the film thickness is by ~ 1 nm steps, corresponding to one quintuple layer (QL) of BiSbTeSe$_2$, which implies that the film growth was in the layer-by-layer mode. Thus, the thickness of BiSbTeSe$_2$ films could be systematically controlled by the growth time (Also, see extended Data Fig. 3). The microstructural properties of the system can be investigated by transmission electron microscopy (TEM). As shown in Fig. 1g, cross-sectional high-angle annular dark-field scanning transmission electron microscopy (HAADF-STEM) with electron energy loss spectroscopy (EELS) reveals that BiSbTeSe$_2$ QLs are periodically stacked on graphene with van der Waals gap spacing and no atomic mixing between layers. The interlayer distance between BiSbTeSe$_2$ and graphene is estimated to be ~ 3.89 Å, while the thickness of a single QL of BiSbTeSe$_2$ is estimated to be 9.97 Å (see Supplementary Information section 4). Additionally, X-ray diffraction detected distinct (110) and (00*l*) peaks for BiSbTeSe$_2$, with lattice parameters *a* = 4.18 Å and *c* = 29.93 Å in the hexagonal unit cell (see Extended Data Fig.2). Energy dispersive X-ray spectroscopy (EDX) of a 30 nm film showed a composition of Bi$_{1.00}$Sb$_{1.05}$Te$_{0.99}$Se$_{1.97}$, closely matching the source material composition of BiSbTeSe$_2$. EDX mapping confirmed the uniform distribution of Bi, Sb, Te, and Se in the film (see Method and Extended Data Fig.4). These results present the capability to epitaxially grow BiSbTeSe$_2$ films with controllable thickness and composition on graphene.

**Local and non-local transport of 9 nm BiSbTeSe$_2$/graphene van der Waals heterostructures**

To investigate the electronic states resulting from the interplay between the gapped 3D-TI surface and the Kekulé graphene in the hybrid system, we performed electrical transport measurements using dual-gate six-terminal Hall bar devices with a BiSbTeSe$_2$/graphene heterostructure sandwiched between h-BN layers and varying the BiSbTeSe$_2$ thicknesses (Figure 2a,b). Since the SOI is expected to be enhanced in thicker BiSbTeSe$_2$ films, we first examined the device with a 9 nm BiSbTeSe$_2$ film, which is anticipated to provide the strongest SOI while preserving the hybridization gap [22]. We employed a dry transfer technique [36] to assemble the BiSbTeSe$_2$/graphene heterostructure and the h-BN films, ensuring a contamination-free process. The detailed process is shown in Method and Extended Data Fig.5. The bottom h-BN layer serves as an atomically flat insulator with minimal surface impurities, while the top h-BN layer protects the BiSbTeSe$_2$ films from oxidation.

Figure 2c presents a color plot of the longitudinal resistance $R_{xx}$ of the dual-gated device for 9 nm BiSbTeSe$_2$ film at $T$ = 0.3 K. $R_{xx}$ is plotted as functions of both top and bottom gate voltages ($V_{TG}$ and $V_{BG}$). By adjusting $V_{TG}$ and $V_{BG}$, the Fermi level of the device is tuned into the band gap, showing a maxima value of $R_{xx}$ ~ 13 kΩ (~ $h/2e^2$) within the range of 0 V ≲ $V_{TG}$ ≲ 0.75 V, 4 V ≲ $V_{BG}$ ≲ 10 V. Fixing $V_{TG}$ ~ 0.25 V and sweeping $V_{BG}$, we observed the plateaus in the resistance of $R_{xx}$ at apparently quantized at a half of the quantum resistance $R_Q$ ($R_Q \sim h/e^2$ or 25.8 kΩ). This value matches the four-terminal resistance $R_{ij;kl} = V_{kl}/I_{ij}$ (where $I_{ij}$ is the current injected from the terminal i to j and $V_{kl}$ is the voltage difference between the terminals *k* and *l*) of a six-terminal device with perfectly transmitting helical edge channels, when measured between any pair of adjacent lateral terminals ($k = l + 1$) [37]. To confirm the presence of helical edge channels in our device, we measured an additional sample with a different length between terminals 2 and 3 ($L \sim$ 4μm) and performed 4-terminal transport measurements under various current and voltage electrode configurations, commonly referred to as non-local transport measurements. Figures 2d–2i present the results of non-local transport measurements for various configurations in two devices with different channel lengths ($L \sim$ 2 and 4μm). We measured $R_{14;14}$, $R_{12;65}$, $R_{13;13}$, and $R_{13;64}$ as functions of the back gate voltage while keeping the top gate voltage fixed. Plateaus were observed in both devices, irrespective of their length differences, with values approximately matching $3R_Q/2$, $R_Q/6$, $4R_Q/3$, and $2R_Q/3$ respectively. Nonlocal transport measurements, interpreted using the Landauer-Büttiker formalism [37], provide experimental evidence for helical edge states, as previously reported [23, 24, 28, 37-41] where $R_{ij;kl}$ is quantized at fractions of the quantum resistance $R_Q$. Unlike trivial edge channels, the resistance values of helical edge channels depend only on the number of intervening voltage probes and are independent of the channel length, as the helical edge states are protected from backscattering by the nontrivial inversion gap and time-reversal symmetry [9]. The observed resistance values closely match with theoretical predictions for helical edge states and remain largely unaffected by channel length, supporting the existence of helical edge channels in our devices. Small deviations of $R_{ij;kl}$ from the ideal quantized values may arise from several factors, including disorder at the contacts [42,43] (see Supplementary information section 1), elastic spin-flip backscattering [44,45] (Supplementary information section 2) and inelastic spin-flip backscattering process via electron-electron interaction [46].

**Temperature and BiSbTeSe$_2$ thickness dependence**

In our heterostructure, the 2D-TI states emerge due to proximity-induced SOI in Kekulé graphene at its interface with the 3D-TI. This interaction positions the Fermi level within the band gaps of both Kekulé graphene and the hybridization surface gap of BiSbTeSe$_2$, while the SOI strength is modulated by the 3D-TI thickness. To examine how

BiSbTeSe$_2$ thickness influences electronic state control, we measured the temperature dependence of $R_{14;23}$ of different thicknesses of BiSbTeSe$_2$ layers (2, 3, 4, 9, and 20 nm), as shown in Figs. 3a–3d and the Extended data Fig. 6.

Figure 3a shows that the temperature-dependent $R_{14;23}$ for the 2 nm device exhibits behavior typical of a trivial insulator. As seen in Extended Data Figure 6a, the Arrhenius plot reveals a thermal excitation gap of $\Delta_1$ = 100 meV in the 200 - 300 K range. Below this temperature range, a crossover occurs to a smaller excitation gap of $\Delta_2$ = 7 meV from 50 K to 150 K, and further to $\Delta_3$ = 0.3 meV between 0.3 K and 20 K. The hybridization gap of 2 nm BiSbTeSe$_2$ can reach up to 100 meV [21, 22], corresponding to $\Delta_1$ = 100 meV, while $\Delta_2$ = 7 meV is attributed to the Kekulé gap in graphene, as supported by the DFT calculations for the BiSbTeSe$_2$/graphene heterostructure discussed in the next section and the Supplementary information. The low-temperature gap $\Delta_3$ is associated with variable-range hopping (VRH), described by $R(T) \propto \exp\left(\frac{T_0}{T}\right)^{\frac{1}{3}}$ as shown in Extended Data Figure 6b. The $n$ = 1/3 exponent is characteristic of 2D VRH transport by localized impurity states within an energy gap [47]. This result suggests that the BiSbTeSe$_2$ film was successfully grown on monolayer graphene, forming the intended structure that enables tuning graphene into the insulating Kekulé state. The extracted hopping temperature $T_0$ = 0.7 K (Extended Data Fig. 6b) further supports the presence of such localized states, which may lead to charge puddle formation [48].

Figure 3b presents the temperature-dependent $R_{14;23}$ for the 3 nm device, revealing significant differences from the insulating behavior observed in the 2 nm device. The $T$-$R_{14;23}$ curve can be divided into three distinct regions; from 0.3 K to 200 K, it remains nearly constant at approximately $R_Q/2$; between 200 K and 250 K, it increases; and above 250 K, it gradually saturates. For the 9 nm device, the $T$-$R_{14;23}$ curve exhibits complex behavior, as shown in Figure 3c. From 0.3 K to 80 K, $R_{14;23}$ remains nearly constant at approximately $R_Q/2$, similar to the trend observed in the 3 nm device. Between 80 K and 140 K, it increases rapidly, followed by a gradual saturation from 140 K to 180 K. Above 180 K, it drops sharply to 300 K. The nearly temperature-independent resistance at ~$R_Q/2$ suggests the presence of helical edge channels in 3 nm and 9 nm devices. The narrower temperature range in the 9 nm device can be attributed to the reduced hybridization gap in the 3D-TI surface states as the BiSbTeSe$_2$ thickness increases [22].

**Interpretations of transport mechanisms**

In a symmetric Hall bar device, where contact-induced disorders have been eliminated, the interplay between random edge defects - arising from the sample fabrication process - and electron-electron (e-e) interactions gives rise to local magnetic moment that can induce elastic spin-flip backscattering, potentially accounting for the slight deviation of the measured resistance from the expected quantized resistance at low temperatures [44] (See Supplementary information). Meanwhile, electron-electron interactions within bulk charge puddles, mediated by electron tunneling to the helical edge states, are widely considered the primary mechanism of inelastic spin-flip backscattering, which leads to the temperature dependence of the measured resistance [46]. These effects contribute to the system's internal resistance ($R_{int}$). In general, $R_{int} = R_Q(\frac{1}{T_{trans}} - 1)$, with $T_{trans}$ is the transmission between two Ohmic contacts, $T_{trans}$ = 1 for the case of the perfect transmission. For a symmetric Hall bar device where the transmission between all adjacent electrodes is constant at $T_{trans}$, the measured $R_{14;23}$ simplifies to: $R_{14;23} = \frac{R_Q}{2}\frac{1}{T_{trans}}$.

By extracting $R_{int}$ using the relation $R_{int} = 2R_{14;23} - R_Q$, the 3,4 and 9 nm devices exhibit power-law behavior with temperature $R_{int} \sim T^\alpha$, as shown in Extended Data Figs 6 c-f. For 9 nm device, $R_{int}$ follows a power-law scaling $R_{int} \sim T^3$ in the low-temperature range of 0.3 K to 80 K. This behavior is consistent with the mechanism proposed in Ref. 46 which attributes the $T^3$ dependence to inelastic spin-flip backscattering induced by charge puddles near the edge. In contrast, such behavior is not observed in thinner BiSbTeSe$_2$ devices 3 and 4 nm) over the same temperature range, likely due to their larger hybridization gap. The rapid increase of $R_{int}$ with power-law behavior for 3 nm and 4 nm device above 200 K (α ~ 5 – 8) and for 9 nm device within the temperature range of 80 K to 140 K (α ~13) may result from a rise in the number of charge puddles with increasing temperature, leading to enhanced backscattering probability. This effect can be attributed to the suppression of electron transfer at the interface of graphene-BiSbTeSe$_2$ due to the mismatch in thermal expansion coefficients—graphene (work function ~4.6 eV) has a negative coefficient [49, 50] while BiSbTeSe$_2$ (work function ~5.2 eV)

[51] has a positive coefficient [52]—which shifts the valence band of BiSbTeSe$_2$ surface states closer to $E_F$ [21, 53]. Considering disorder-induced mobility edge formation at the band edge of gapped 3D-TI surface states [54], this energy shift generates a few charge puddles near the helical edge states and induces the increasing of the internal resistance above 200 K for 3 nm and 4 nm device with a large hybridization gap of BiSbTeSe$_2$. For the 9 nm device with a small hybridization gap, a few charge puddles exist at 0.3 K and this energy shift leads to an increase in the number of charge puddles near the helical edge. Thereby backscattering probability of helical edge enhances above 80 K, resulting in the rapid increase of $R_{int}$ with power-law behavior.

Above 140 K, the enhancing backscattering probability likely suppresses the helical-edge contribution to the overall electrical resistance. Instead, thermally excited carriers, from the mobility edge into localized 3D-TI surface states, may dominate the $R_{int}$ (Fig. 3e). Indeed, the saturation of $R_{int}$ at approximately ~0.7 h/e² from 140 K to 180 K closely approximates the electrical resistance saturation ~0.8 h/e² of a 10 nm-thick BiSbTeSe$_2$ film, where a small hybridization gap is formed in the 3D-TI surface states [22]. At higher temperatures (180 K to 200 K) for 9 nm device, when the mobility edge of BiSbTeSe$_2$ surface states reaches $E_F$, resistance drops sharply. Beyond 200 K for 9 nm device, transport is dominated by surface and bulk carriers in BiSbTeSe$_2$. The Arrhenius plot for this regime yields an excitation gap of Δ = 21 meV, consistent with the value extracted from the bulk electrical resistance of BiSbTeSe$_2$ crystals [22] (Extended Data Figure 6g).

For the 20 nm device as shown in Fig. 3d, $R_{14;23}$ remains nearly constant at approximately ~ 0.4 h/e² from 0.3 K to 150 K and then decreases with increasing temperature. In this device, although the enhanced SOI is expected to increase the inversion gap of graphene, the hybridization gap of BiSbTeSe$_2$ closes [22], resulting in electrical resistance dominated by 3D-TI surface states from 0.3 K to 150 K. In fact, electrical resistance saturation to ~ 0.4 h/e² has been observed in BiSbTeSe$_2$ films with a thickness of 14 nm where the 3D-TI surface gap is closed [22], quantitatively consistent with our 20 nm device results. Above 150 K, thermally excited bulk holes of BiSbTeSe$_2$ suppress the $R_{14;23}$. The Arrhenius plot estimates an excitation gap of 33 meV, which may correspond to thermal excitation from the bulk or metallic surface states to impurity states

## Confirmation of the existence of helical edge states in vdW BiSbTeSe$_2$/graphene by DFT calculations

To investigate the mechanisms responsible for the emergence of a 2D-TI phase in the BiSbTeSe$_2$/graphene vdW heterostructures, we performed ab initio calculations within the framework of density functional theory (DFT). The heterostructures were modeled by placing 2QL and 3QL (1QL ~ 1nm) BiSbTeSe$_2$ slabs on monolayer graphene. A relaxed interlayer spacing d ~ 3.5Å was used, with the quintuple-layer (QL) stacking order from the bottom as Te-Sb-Se-Bi-Se. In this configuration, the Te atoms are centrally aligned with the carbon hexagonal rings, following a $\sqrt{3}a \times \sqrt{3}a$ stacking pattern, where $a$ is the graphene lattice constant (Supercell of graphene ~ 4.23 Å), as shown in Fig.4a. The band structure calculations in Figs. 4 b and 4c reveal that, in this configuration, the graphene DC is folded to the Brillouin zone center [15, 16, 18], aligning it at the Fermi level and opening a gap of 12.8 meV and 15.6 meV for 2QL and 3QL due to the enhancement of intrinsic SOI in graphene.

To confirm the presence of a 2D-TI phase in our system, we calculated the Z$_2$ topological index using the Fu-Kane criterion [55]. For simplicity, the calculation was performed on a sandwich structure of BiSbTeSe$_2$/graphene/BiSbTeSe$_2$ with the interfacial stacking order Te-Sb-Se-Bi-Se, incorporating parameters consistent with the experimental system. Upon inclusion of SOI, the parity eigenvalues at time-reversal invariant momentum (TRIM) points yield a nontrivial Z$_2$ index (Z$_2$ = 1), indicating that SOI induces band inversion, as shown in Fig. 4c.

## Outlook

Our results demonstrate that the BiSbTeSe$_2$/graphene vdW heterostructure effectively transforms the massless Dirac states in graphene into high-temperature 2D-TI states. Precise tuning of the Fermi levels in both BiSbTeSe$_2$ and graphene near their DPs is crucial for engineering new topological phases through a combination of interfacial Kekulé distortion and proximity-induced SOI via band hybridization. Our straightforward growth method enables fabrication of 2D-TI systems exhibiting robust helical edge states up to 200 K, without the need for extreme conditions such as high pressure or strong magnetic field. By controlling the BiSbTeSe$_2$ thickness, it is possible to optimize both the proximity-enhanced SOI in graphene and the suppression of the hybridization gap on the 3D-TI surface. This tunability opens the prospect of realizing helical edge states even above room temperature. High-temperature 2D-TIs are particularly promising for spintronic applications. The combination of helical edge states with ferromagnetic materials is expected to yield spin-to-charge conversion efficiencies 100–1000 times greater than those of conventional 3D-TIs and Rashba conductors via the inverse Edelstein effect. Such enhancements enable efficient generation and detection of spin currents, together with ballistic, long-range spin transport—critical features for the design of next-generation spintronic devices [56,57]. Furthermore, these systems offer a new platform for topological quantum technologies. In particular, coupling high-temperature 2D-TIs to superconductors could allow the observation and control of Majorana fermions through the superconducting proximity effect [58,59]

## Acknowledgements
Y.T. acknowledges support from the Japan Society for the Promotion of Science (JSPS) under Grant Nos. JP22K04867, 19K05195. N. H. T. acknowledges the support from JSPS under Grant No. 22K13989 and RIKEN Incentive Research Projects (202001062017). M.-C. J. thanks for the support from the RIKEN IPA program and the computing time in RIKEN Hokusai BigWaterfall2. M.-C.J. and G.-Y.G. acknowledge the support from the Ministry of Science and Technology and the National Center for Theoretical Sciences (NCTS) of the R.O.C. R. A. thanks for the support by the RIKEN TRIP initiative (RIKEN Quantum, Advanced General Intelligence for Science Program, Many-body Electron Systems) and the support from JSPS under Grant No. JP25H01252.

## Author information
These authors contributed equally: Yoichi Tanabe and Ngoc Han Tu
### Authors and Affiliations
**Department of Applied Science, Graduate School of Science, Okayama University of Science, Okayama, Japan**
Yoichi Tanabe
**RIKEN Center for Emergent Matter Science (CEMS), Saitama, Japan**
Ngoc Han Tu, Ming-Chun Jiang, Yi Ling Chiew, Ryo Ito, Yuya Shimazaki, Kiyohiro

Adachi, Daisuke Hashizume, Xiuzhen Yu, Ryotaro Arita and Michihisa Yamamoto

**Department of Physics and Center for Theoretical Physics, National Taiwan University, Taipei, Taiwan**

Ming-Chun Jiang, Guang-Yu Guo

**Institute for Chemical Research (ICR), Kyoto University, Kyoto, Japan**

Mitsutaka Haruta

**Quantum-Phase Electronics Center and Department of Applied Physics, The University of Tokyo, Tokyo, Japan**

Yuya Shimazaki, David Pomaranski and Michihisa Yamamoto

**Physics Division, National Center for Theoretical Sciences, Taipei 10617, Taiwan**

Guang-Yu Guo

**Department of Physics, The University of Tokyo, Tokyo, Japan**

Ryotaro Arita

**Author Contributions**

Y.T. grew the $BiSbTeSe_2$ thin film on mono layer graphene substrates. N.H.T. characterized $BiSbTeSe_2$ thin film by AFM and EDX. N.H.T. fabricated dual gate field effect transistors and performed electrical transport measurements. R.I, D. P. and N. H. T. set up for the electrical transport measurement. Y. S. and N. H. T. prepared for the dry transfer system. Y. L. C., M. H. and X. Y. performed the TEM measurement. K.A performed the X-ray measurement. K. A. and D. H performed the X-ray diffraction analysis. M.-C. J., G.-Y. G., and R.A. carried out theoretical calculations. Y.T., N.H.T., and M.C.J. analyzed data. Y.T, N. H. T. M.C. J. and M. Y. wrote the manuscript. All authors discussed the results and commented on the manuscript. Y.T. and N.H.T. conceived and designed the project under the supervision of M.Y.

**Corresponding Authors**
Corresponding to Yoichi Tanabe, Ngoc Han Tu, and Michihisa Yamamoto

**Ethics declarations**
Competing interests

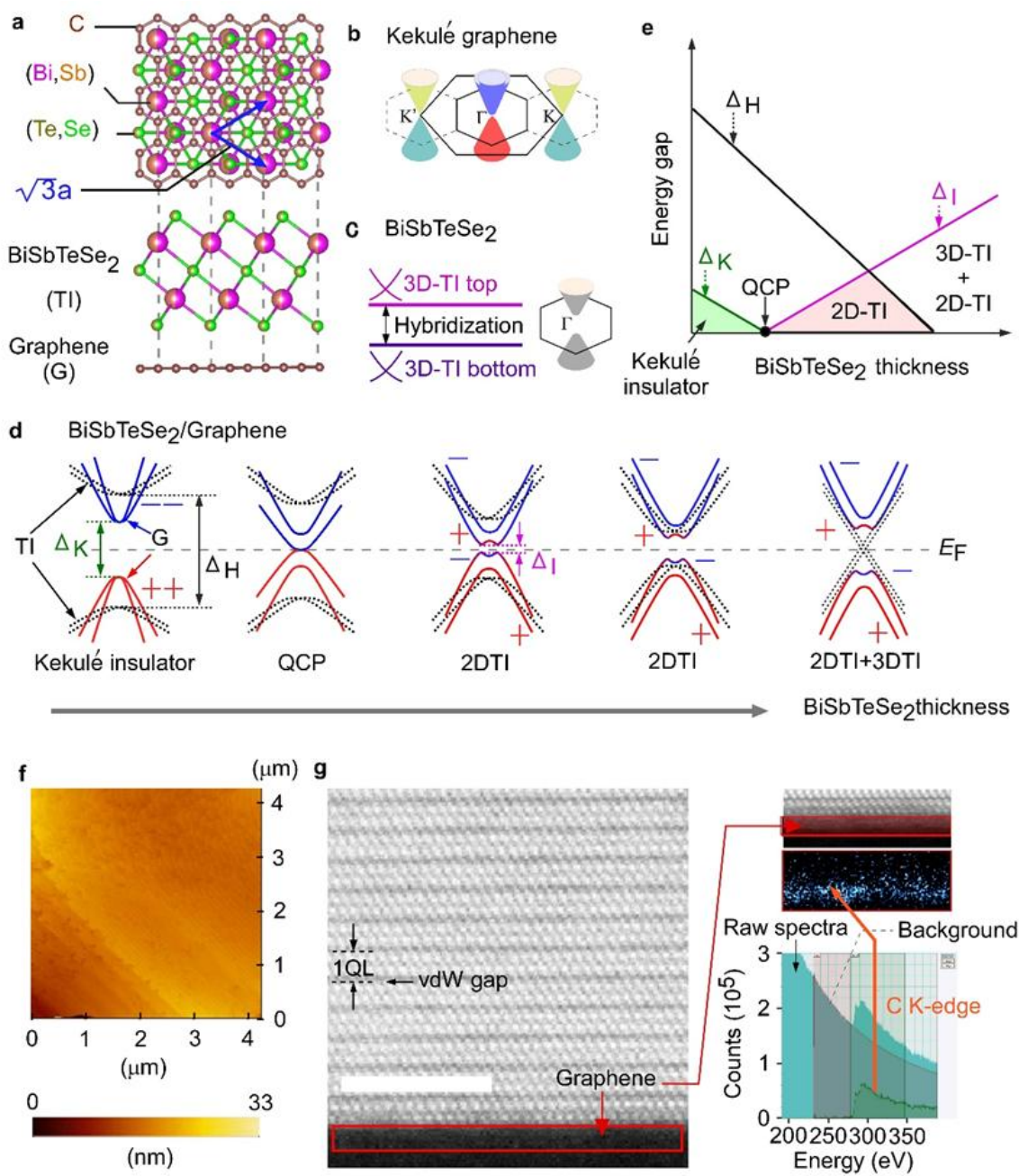

**Figure 1. Fabrication of BiSbTeSe$_2$/graphene van der Waals heterostructures and the underlying mechanism for inducing 2D topological insulator states in the system.** a. Schematic of the stacking of BiSbTeSe$_2$ and graphene in the heterostructure film from the top and side view. The blue arrows indicate the in-plane translation vectors

in the hexagonal unit cell of BiSbTeSe$_2$ with the graphene lattice constant $a$ as the unit. b-e. Schematic illustration of the proposed electronic states in the system. In ultrathin BiSbTeSe$_2$ films, a hybridization gap emerges due to the coupling between the top and bottom topological surface states, while in graphene, a gap is opened by the Kekulé distortion. As the thickness of the BiSbTeSe$_2$ film increases, the spin–orbit interaction (SOI) is enhanced, driving a transition from a trivial insulator to a 2D topological insulator (2D-TI). f. AFM image of the system with 30 nm BiSbTeSe$_2$ grown on graphene, showing step heights of approximately 1 nm—corresponding to one quintuple layer (1QL)—which indicates a layer-by-layer growth mode. g. Cross-sectional HAADF-STEM image of the system, along with EELS spectra of the BiSbTeSe$_2$/graphene vdW heterostructure, confirming that BiSbTeSe$_2$ films can be successfully grown on monolayer graphene. The scale bar is 5 nm.

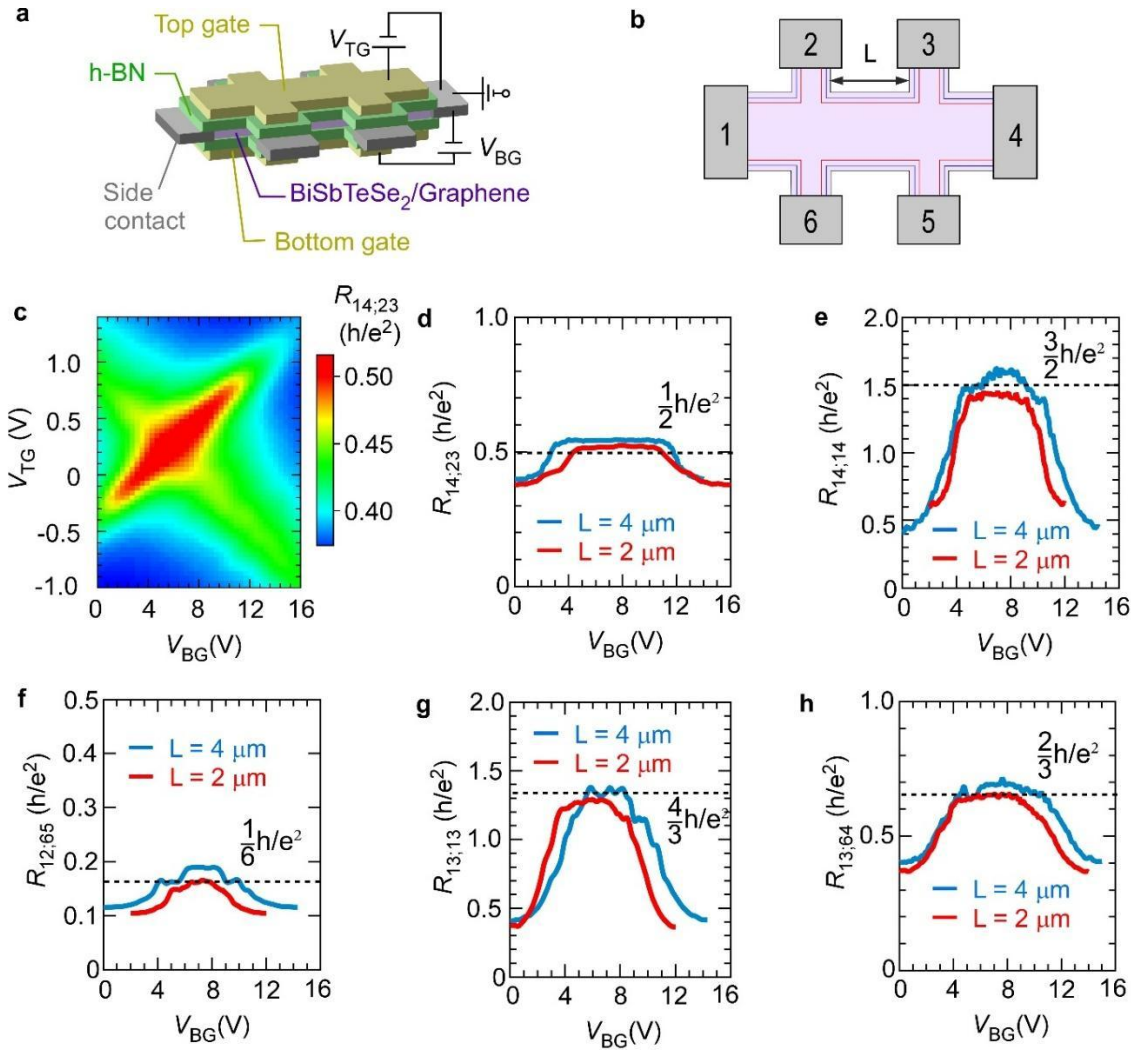

**Figure 2. Schematic of devices and non-local transport measurements of BiSbTeSe$_2$/graphene van der Waals heterostructures with a 9 nm of BiSbTeSe$_2$ film at 0.3 K.** a. Schematic image of the dual-gate BiSbTeSe$_2$/graphene heterostructure film field-effect transistor using h-BN as the gate capacitor. b. Top view of the Hall bar structure with the number of electrode terminals. Red and blue lines are the channels with up and down spin. c. Color map of four-terminal resistance $R_{14;23}$ as a function of the top gate voltage, $V_{TG}$, and the bottom gate voltage, $V_{BG}$ for the voltage terminal spacing of 2 µm. d-h. $V_{BG}$ dependence of the 4-terminal resistances for various configurations of the current and voltage terminals for $V_{TG}$ = 0.2 V. The red and blue curves show results for the short-channel device with $L$ = 2 µm and $L$ = 4 µm, respectively ($L$ is the length between contact 2 and 3). Dashed lines indicate the expected quantized resistances for the helical edge state calculated using the Landauer-Büttiker formalism.

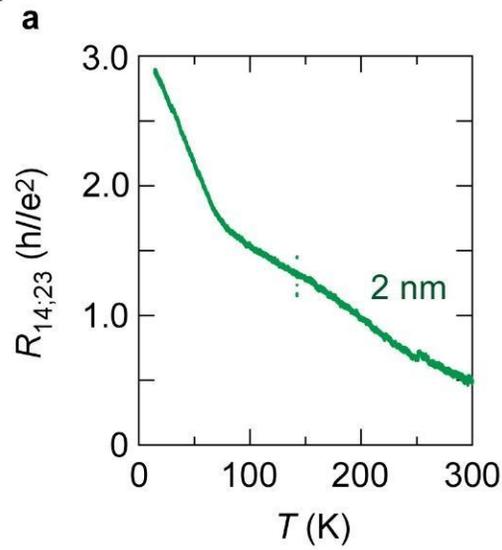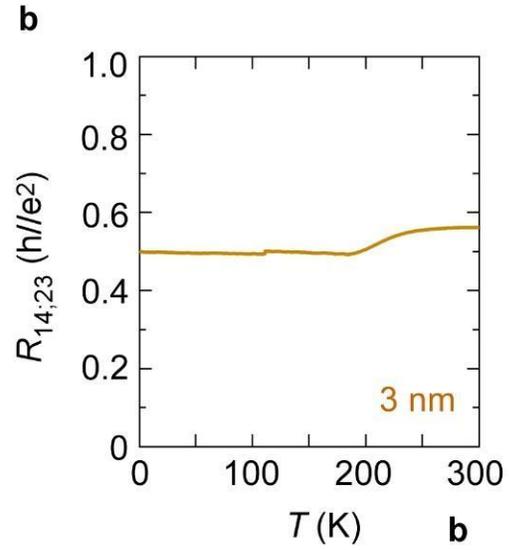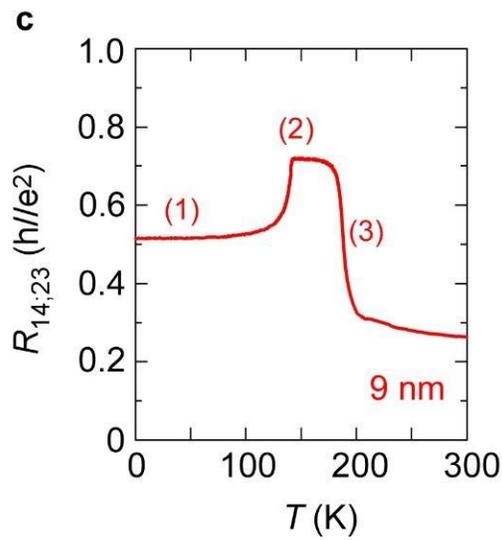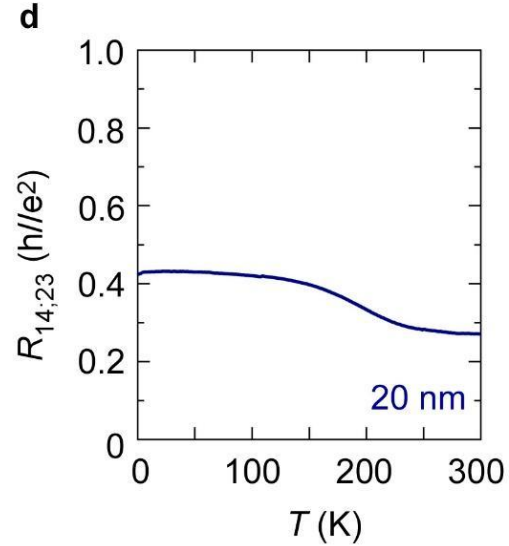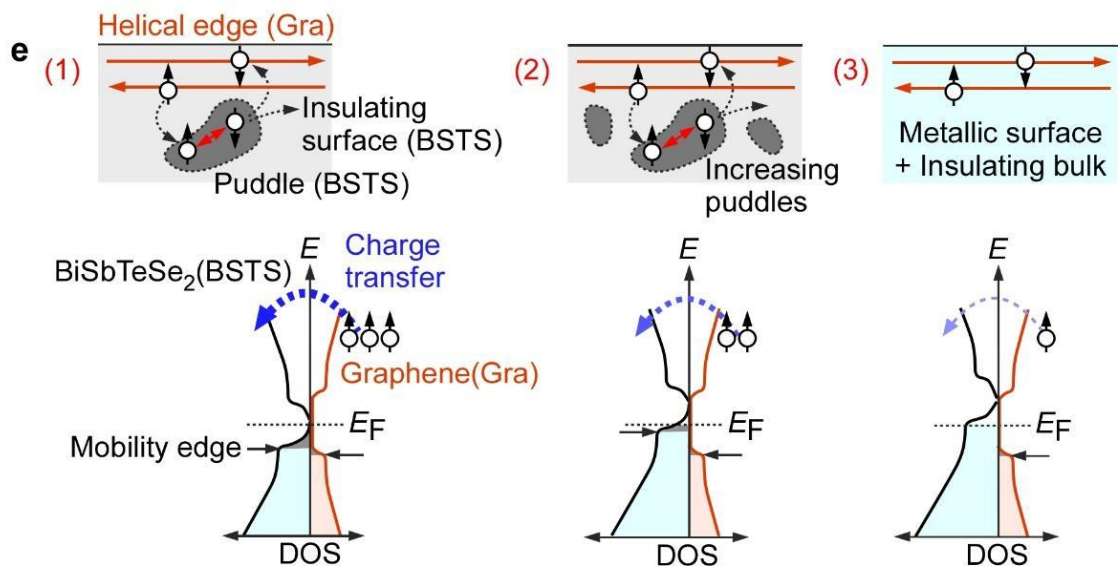

**Figure 3. Temperature dependence of resistance $R_{14;23}$ in BiSbTeSe$_2$/graphene heterostructures with varying BiSbTeSe$_2$ thickness.** a - d, Temperature-dependent resistance $R_{14;23}$ for BiSbTeSe$_2$ thicknesses of 2 nm, 3 nm, 9 nm and 20 nm, respectively e, Schematic illustration of the electronic states and backscattering mechanisms in the heterostructure with a 9nm BiSbTeSe$_2$ layer: (1) $T$ = 0.3 K – 80 K: The hybridization gap of BiSbTeSe$_2$ and the inversion gap in graphene are aligned, forming a 2D-TI phase. Due to the small hybridization gap [22] and emergence of a mobility edge in the gapped 3D-TI surface states [54], metallic charge puddles form near the edge, leading to backscattering and a $T^3$ behavior in $R_{14;23}(T)$. (2) $T$ = 80 K – 180 K: Interfacial electron transfer is suppressed, possibly due to thermal expansion mismatch. Simultaneously, 3D-TI surface-derived charge puddles increase, resulting in a rapid increase in backscattering from 80 K to 140 K. Above 140 K, this enhanced backscattering quenches the helical-edge contribution, and thermally activated metallic holes in the valence band of the BiSbTeSe$_2$ 3D-TI surface states take over conduction, causing $R_{14;23}$ to saturate as the temperature increases. (3) $T$ > 180K: As the mobility edge approaches the Fermi level, metallic contributions from the 3D-TI surface dominate, resulting in a rapid decrease in $R_{14;23}$.

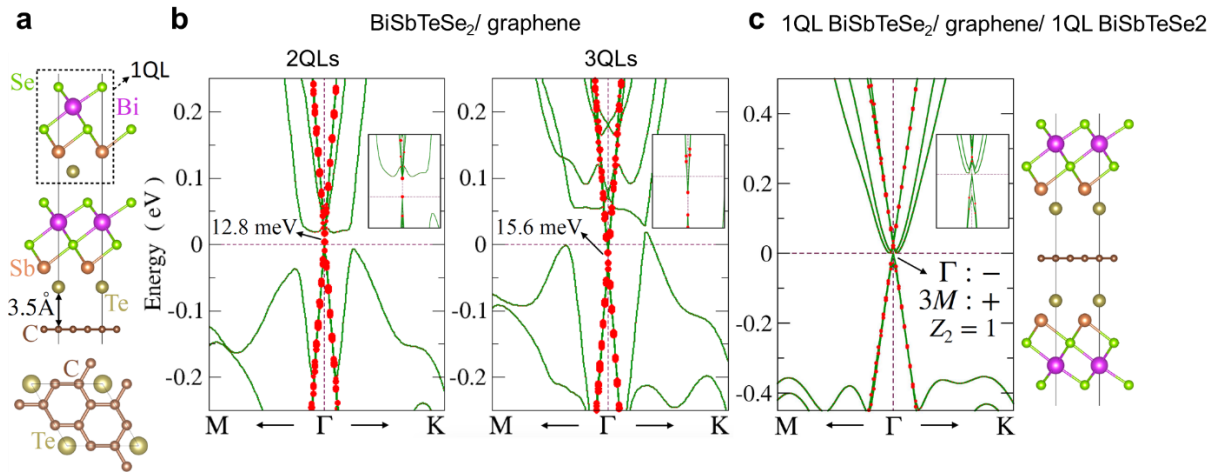

**Figure 4.** The band structure of the graphene-BiSbTeSe$_2$ heterostructures from DFT calculation along *M-Γ-K* direction. a. Side view (top) and top view (bottom) of the crystal structure of BiSbTeSe$_2$ – graphene heterostructure. The interface sequence C–Te–Sb-Se-Bi-Se was chosen to minimize the energy difference between the two DPs. b Band structure of the BiSbTeSe$_2$ / graphene heterostructure with two and three quintuple layers (QLs) of BiSbTeSe$_2$, corresponding to film thicknesses of approximately 2 nm and 3 nm, respectively, in the presence of SOI. The green lines represent the electronic states originating from BiSbTeSe$_2$, while the red dots highlight the graphene- electronic states. Band gaps are observed for both 2QL and 3QL configurations, with sizes of approximately 12.8 meV and 15.6 meV, respectively. The insets show the magnitude views around the gaps. c. Band structure of 1QL BiSbTeSe$_2$/graphene/1QL BiSbTeSe$_2$ sandwich structure. The nontrivial topological invariant $Z_2$ = 1 confirms the topological nature of the system, suggesting that a topological gap can be obtained in BiSbTeSe$_2$/graphene systems.

**Method**

*1. van der Waals epitaxial growth of BiSbTeSe$_2$ film on graphene by physical vapor deposition*

~10 mg of BiSbTeSe$_2$ single crystal pieces were ground on an agate mortar to prepare BiSbTeSe$_2$ powder as the source material. The BiSbTeSe$_2$ powder was introduced through a funnel into the bottom of a 13 cm long quartz tube (inner tube) with an outer diameter of 13 mm, which was sealed at one end. As shown in Extended Data Fig.1, the inner tube was placed inside a 28 mm diameter of quartz tube (outer tube) installed in a three-zone tube furnace such that the source material was placed in the center (upstream) of the furnace, with graphene transferred to a SiO$_2$/Si substrate near the opening end of the inner tube. The inner gas of the outer tube is replaced with argon gas several times and then evacuated to ~10$^{-1}$ Torr using a rotary pump. The temperature was increased until the upstream side of the furnace reached approximately 490 °C and the downstream side reached around 350 °C simultaneously. Once the target temperatures were achieved, BiSbTeSe$_2$ was grown on graphene for 30 seconds to 2 minutes, followed by rapid quenching to room temperature.

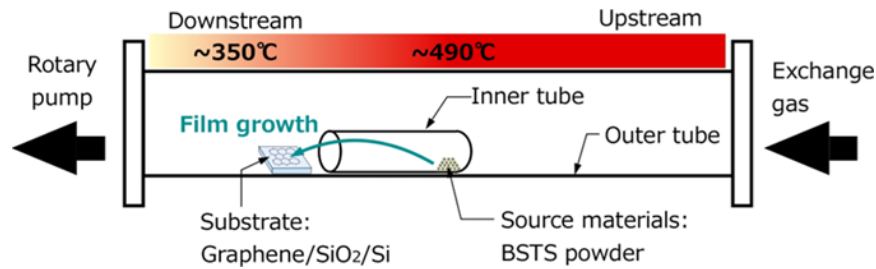

**Extended Data Figure 1|** Schematic illustration of the growth process of BiSbTeSe$_2$/graphene vdW heterostructures via physical vapor deposition.

*2. Characterization of BiSbTeSe$_2$/graphene van der Waals heterostructures*

BiSbTeSe$_2$/graphene van der Waals heterostructures were characterized by out-of-plane and in-plane X-ray diffraction of the thin films, cross-sectional transmission electron microscopy, surface observation by atomic force microscopy (AFM), and Energy dispersive X-ray (EDX), shown in Extended Data Figs. 2-5.

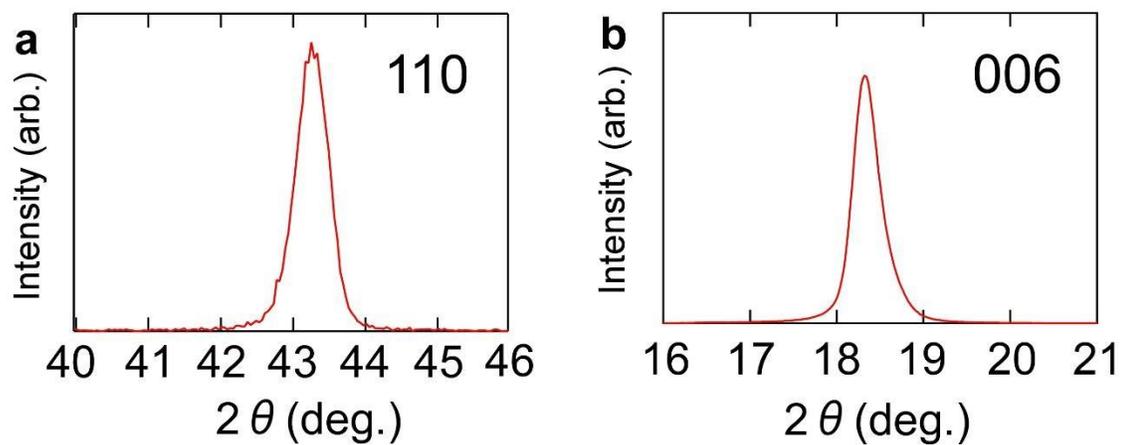

**Extended Data Figure 2|** X-ray diffraction of BiSbTeSe$_2$/graphene vdW heterostructures.

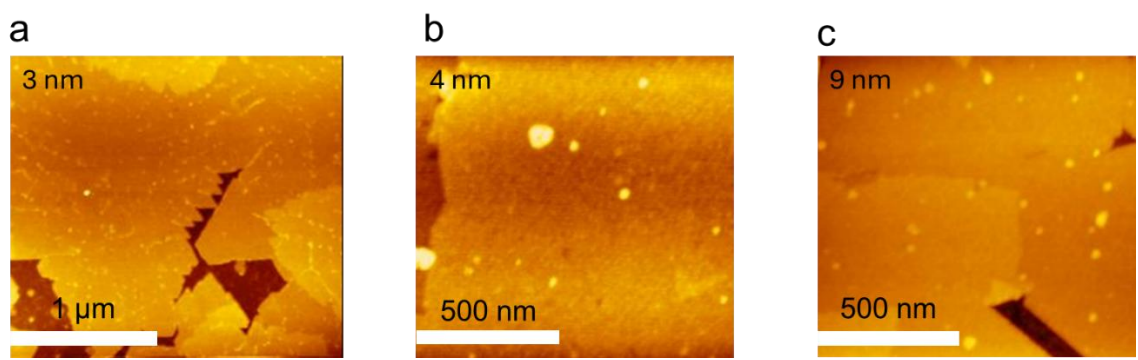

**Extended Data Figure 3|** AFM images with a. 3 nm, b. 4nm, c. 9 nm thickness of BiSbTeSe$_2$ layers.

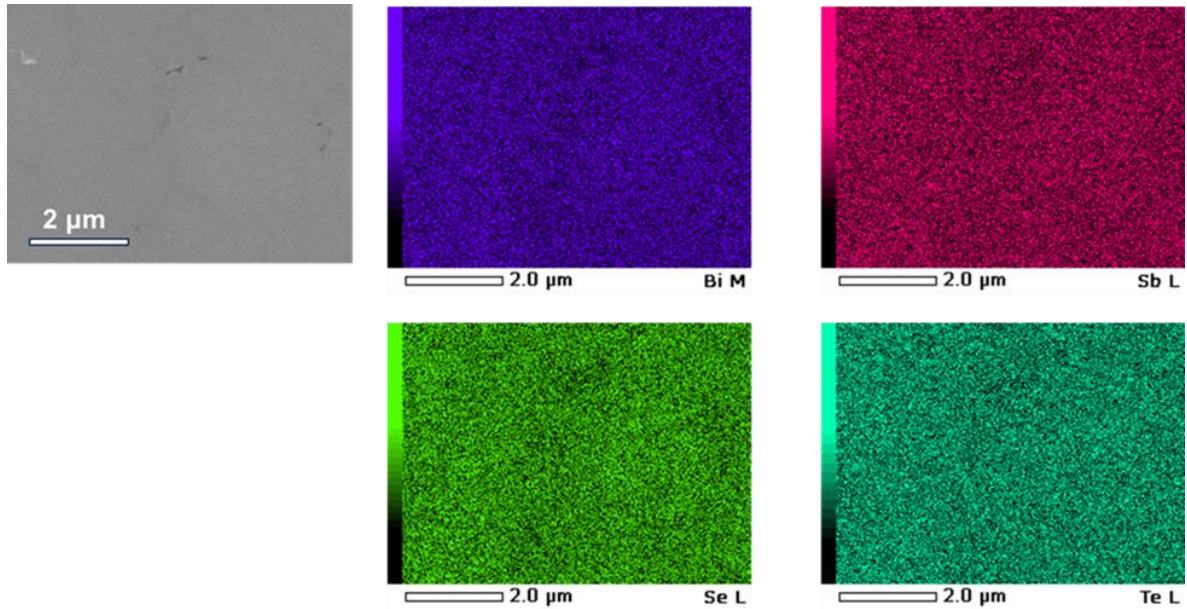

**Extended Data Figure 4|** EDX mapping data of BiSbTeSe$_2$/graphene vdW heterostructures (BiSbTeSe$_2$ thickness is 30 nm, same with the TEM measurement).

## 3. Device fabrication

We use the dry transfer method to fabricate the dual gate devices [36]. First, we prepared a stamp on a glass slide, consisting of a droplet-shape of polydimethylsiloxane (PDMS) gel covered by a thin polypropylene carbonate (PPC) layer. The stamp was baked in an oven at 130°C for 15 minutes and brought into a glovebox. The dry stacking system and the optical microscope were put inside the glove box to facilitate these operations. Inside glovebox, we conducted the assembly of hBN/vdW BiSbTeSe$_2$-graphene/hBN for dual gate devices on the SiO$_2$/Si substrates. We picked up hBN flake and stacked it with the grown graphene/ BiSbTeSe$_2$ heterostructures, then placed them on another substrate with prepared hBN flakes. Heating the structure to 180°C for 5 minutes left the PPC/hBN/BiSbTeSe$_2$/graphene/hBN on the new substrate. Then we removed the PPC layer by keeping 1 hour in chloroform. The Hall bars were etched using an inductively coupled plasma (ICP) etching machine with the mixing of Cl$_2$/SiCl$_4$ gas. The gold top gate and edge contacting electrodes were fabricated by depositing Cr/Au (2/50 nm). For a detailed fabrication, see Extended Data Fig. 5.

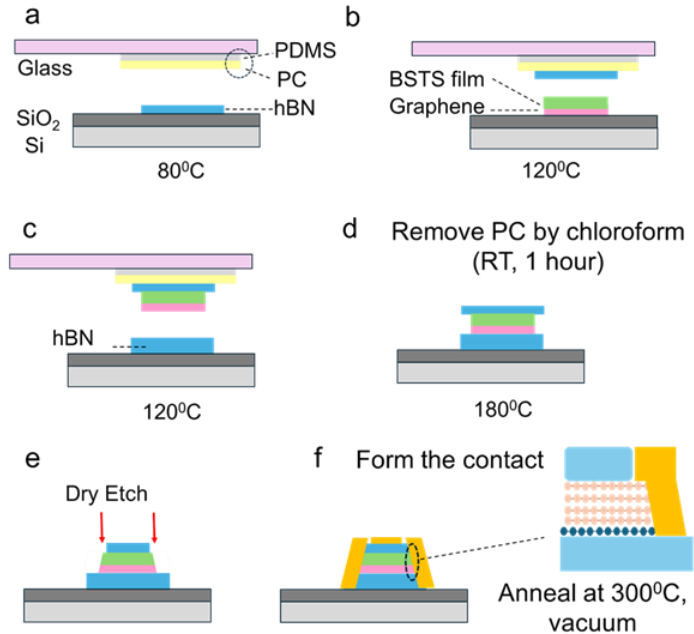

**Extended Data Figure 5|** Fabrication of a dual gate device by dry transfer technique.

## 4. Set up of electrical transport measurements

Low-temperature electrical transport measurements were performed in a helium-3 refrigerator (Niki Glass Company Ltd.) with a base temperature of 0.3 K and magnetic field up to ± 9 T. Four probe resistances of the devices were measured using a NF 5650 lock-in amplifiers operated at a frequency of 13 Hz. The devices were sourced with a constant AC current of 1 nA. Two Keithley 2400 source meters were utilized to source DC gate voltages separately to the top and bottom gate electrodes.

## 5. First-principles calculations of the BiSeTeSe$_2$/graphene vdW heterostructure

First-principles calculations of the BiSeTeSe$_2$/graphene vdW heterostructure are based on the density functional theory (DFT) with the generalized gradient approximation in the form of Perdew-Burke-Ernzerhof [60, 61]. The highly accurate projector-augmented wave (PAW) method [62], as implemented in the Vienna ab initio simulation package (VASP) [63,64], is used. A large plane-wave cutoff energy of 550 eV is used, and a $\Gamma$-centered 12×12×1 k-mesh is adopted for the Brillouin zone integration. The band structures are plotted with the help of VASPKIT [65], and the structures are visualized with

VESTA [66]. The BiSbTeSe$_2$/graphene vdW heterostructure is modeled utilizing the supercell approach with a vacuum layer of about 25 Å thickness. The experimental lattice parameters of BiSeTeSe$_2$ and the experimental atomic positions are used. Two quintuple layers (QLs) of BiSbTeSe$_2$ are set on top of a single graphene layer $\sqrt{3} \times \sqrt{3}$ with the SbTe interface. A $\sqrt{3} \times \sqrt{3}$ commensurate stacking is applied between BiSeTeSe$_2$ and graphene, with the lowest atom of BiSeTeSe$_2$ located in the center of the graphene hexagonal ring. The interlayer distance between BiSeTeSe$_2$ and graphene is calculated via structural optimization along the c direction, in which we have included the DFT-D2 van der Waals correction of Grimme [67]. The optimized interlayer distance is 3.5 Å, which agrees well with the transmission electron microscope (TEM) measurement of 3.9 Å. Note that from the TEM measurements, the atomic composition is the same within each layer of the solid solution of BiSeTeSe$_2$. Therefore, a supercell along the *a*-axis and *b*-axis is not needed. For the Z$_2$ number determination, we construct graphene sandwiched by 1QL of BSTS with the SbTe interface to retain the inversion symmetry. The parity eigenvalues are determined by the Irvsp program [68].

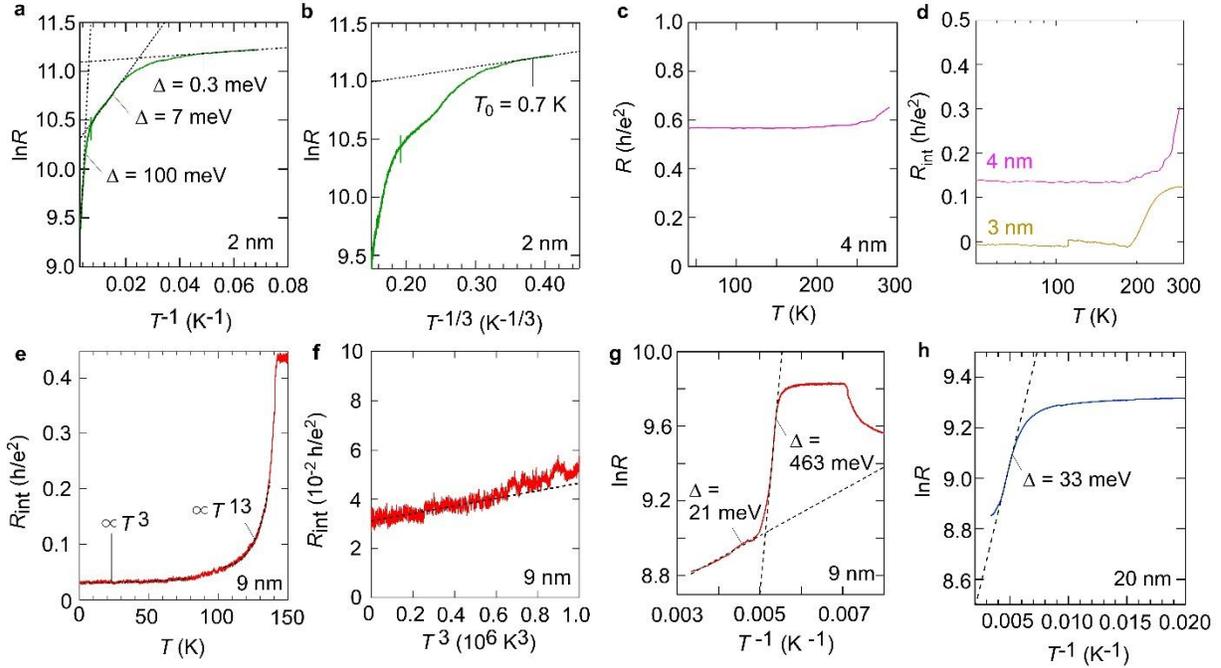

**Extended Data Figure 6|** Detailed analysis of the temperature-dependent $R_{xx}$ ($R_{14;23}$) with different thicknesses of the BiSbTeSe$_2$ film a. Arrhenius plot for 2 nm device. The dashed lines are fitting curves of the data estimating the activation gap Δ in the three different temperature regimes. b. $\log R - T^{-\frac{1}{3}}$ plot of 2D variable range hopping for 2 nm sample. The dashed line is fitting curve of the data estimating characteristic temperature of 2D variable range hopping, $T_0$. c. $R$-$T$ curve for 4 nm device. d. $R_{int}$-$T$ curve for 3 nm and 4 nm device. e. $R_{int}$-$T$ curve for 9 nm sample with a $T^\alpha$ fitting curve. f. $R_{int}$ -$T^3$ curve for 9 nm sample. g. Arrhenius plot for 9 nm sample. The dashed lines are fitting curves of the data estimating the activation gap Δ from 180 K to 200 K and above 200 K. h. Arrhenius plot for 20 nm sample. The dashed lines are fitting curves of the data estimating the activation gap Δ ~ 33 meV.

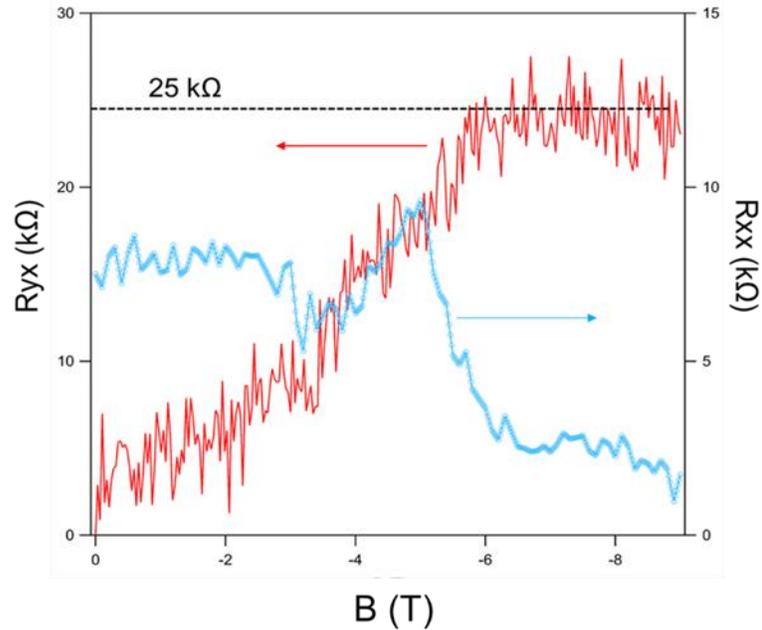

**Extended Data Figure 7|** Observation of the quantum Hall effect at $\nu = 1$ for the thickness of 20 nm BiSbTeSe$_2$ at 0.3K.